\documentclass[12pt,preprint]{aastex}

\shorttitle{[CI] in APM~08279+5255 at z=3.91}
\shortauthors{WAGG ET AL.}

\slugcomment{Accepted by ApJ, July 11, 2006}

\def\ci{{[CI]~$^3$P$_1 - ^3$P$_0$ }}
\def\cup{{[CI]~$^3$P$_2 - ^3$P$_1$ }}
\def\water{{H$_2$O~1$_{10} - $1$_{01}$ }}

\begin{document}

\title{Atomic Carbon in APM~08279+5255 at z=3.91 
\footnote{Based on observations carried out with the IRAM Plateau de
 Bure Interferometer. IRAM is supported by INSU/CNRS (France), MPG (Germany) 
and IGN (Spain).}}

\author{
J. Wagg,$^{2,3}$ D.J. Wilner,$^{2}$ 
R. Neri,$^{4}$ D. Downes,$^{4}$ and T. Wiklind,$^{5}$
}

\affil{$^{2}$Harvard-Smithsonian Center for Astrophysics, 
             Cambridge, MA, 02138}
\email{jwagg@cfa.harvard.edu}

\affil{$^{3}$Instituto Nacional de Astrof\'isica, \'Optica y Electr\'onica 
             (INAOE), Aptdo. Postal 51 y 216, Puebla, Mexico}

\affil{$^{4}$Institut de Radio Astronomie Millim\'etrique, 
            St. Martin d'H\`eres, F-38406, France}

\affil{$^{5}$ESA Space Telescope Division, STScI, 
             3700 San Martin Drive, Baltimore, MD 21218, USA}

\begin{abstract}
We present a detection of \ci emission in the lensed quasar 
APM~08279+5255 at z=3.91 using the IRAM Plateau de Bure interferometer. 
The [CI] line velocity and width are similar to the values of 
previously detected high-J CO and HCN lines in this source,
suggesting that the emission from all of these species arises from 
the same region. The apparent luminosity of the [CI] line is
$L'_{\rm CI} = (3.1 \pm 0.4) \times 10^{10}$~K~km~s$^{-1}$~pc$^2$,
which implies a neutral carbon mass, 
$M_{\rm CI} = (4.4 \pm 0.6)m^{-1} \times10^{7}$~M$_{\odot}$, 
where $m$ is the lensing magnification factor.
The [CI] line luminosity is consistent with the large molecular gas mass 
inferred from the nuclear CO line luminosity 
($\sim$10$^{11} m^{-1}$~M$_{\odot}$).
We also present an upper limit on the \water line luminosity in 
APM~08279+5255 of, 
$L'_{\rm H_2O} < 1.8 \times 10^{10}$~K~km~s$^{-1}$~pc$^2$ (3-$\sigma$).
\end{abstract}


\section{Introduction}

The ultraluminous quasar APM~08279+5255 at z=3.91 (Irwin et al.\ 1998) 
is strongly gravitationally lensed (Ledoux et al.\ 1998; Ibata et al.\ 1999; 
Egami et al.\ 2000) and appears to be one of the most luminous objects 
in the Universe, with an apparent infrared luminosity 
of $\sim$10$^{15}$~L$_{\odot}$ (Lewis et al.\ 1998). 
Observations over a wide range of wavelengths 
(e.g., Downes et al.\ 1999;
Ellison et al.\ 1999; 
Gallagher et al.\ 2002;
Soifer et al.\ 2004;
Wagg et al.\ 2005) have shown that APM08279+5255 contains 
an active nucleus and likely a starburst component.

The combination of extreme intrinsic luminosity and amplification by
strong gravitational lensing 
has allowed APM~08279+5255 to be detected in many rotational lines 
of CO (J=4-3 and J=9-8 Downes et al.\ 1999, hereafter D99;
J=1-0 and J=2-1 Papadopoulos et al.\ 2001; Lewis et al.\ 2002 ), 
including the unique 
detection in the high excitation J=11-10 line 
(A. Weiss et al., in preparation). 
From high angular resolution observations, D99 suggest that the
high-J CO emission arises in a warm, dense circumnuclear disk 
of sub-kiloparsec size. The detection of both HCN J=5-4 
and HCO$^+$ J=5-4 line emission support the idea that the molecular 
gas is dense (Wagg et al.\ 2005; Garcia-Burillo et al.\ 2006)

Given the wealth of molecular lines detected in APM~08279+5255, 
observations of additional diagnostic lines can provide further
constraints on the physical conditions in this high-redshift object.
One important tracer of the dense neutral gas within a galaxy's
interstellar medium is atomic carbon, in particular the \ci line
with rest frequency 492.161~GHz (G\'erin \& Phillips 1998, 2000). 
This line has not been 
observed widely in local galaxies due to poor atmospheric transmission 
near this frequency; however, this line is redshifted into better
atmospheric windows for objects at redshifts, z~$\ga$~1.  
The \ci line provides a complimentary probe of the physical conditions 
of dense, neutral gas, and the [CI]/CO ratio may also provide 
information on the role of X-rays in the molecular excitation 
(Maloney et al. 1996).
In addition, Papadopoulos et al.\ (2004a, b) argue that this line 
can provide a measure of the total molecular hydrogen (H$_2$) gas mass, 
independent of CO line emission.

To date, [CI] emission has been detected in four objects at z~$>2$, 
most of which are thought to be gravitationally lensed. 
The $^3$P$_1 - ^3$P$_0$ 
line was first detected in H1413+117 at z=2.6 (the ``Cloverleaf''; 
Barvainis et al.\ 1997), and more recently the higher excitation
$^3$P$_2 - ^3$P$_1$ line was also detected in this object (Weiss et al.\ 2003).
Other high-redshift objects with \ci line detections include
IRAS~F10214 at z=2.3 (and an upper-limit on the \cup line intensity 
is reported by Papadopoulos 2005), SMM~J14011+0252 at z=2.6 
(Weiss et al.\ 2005), and PSS~2322+1944 at z=4.1 (Pety et al.\ 2005).

Another important species that acts as a major coolant of dense gas
is water. In general, H$_2$O lines at local velocites are not accessible 
from the ground, but they have been observed from space and are ubiquitous 
in dense molecular cloud cores in the Galaxy 
(Ashby et al.\ 2000; Snell et al.\ 2000). 
At high-redshift, a tentative detection of the H$_2$O~2$_{11} - $2$_{02}$ 
line, with rest frequency 752.033~GHz, was reported in IRAS~F10214 at z=2.3 
(Encrenaz et al.\ 1993; Casoli et al.\ 1994). At more modest redshifts, 
the fundamental transition of ortho-water, \water has been detected in
 absorption towards B0218+357 at z=0.685 (Combes \& Wiklind 1997).
For APM~08279+5255, the lower excitation \water line with rest frequency 
556.936~GHz is redshifted into the 3~millimeter atmospheric window. 

Here we report a detection of \ci line emission
in APM~08279+5255 at z=3.91, and an upper-limit on \water line emission, 
obtained with the IRAM Plateau de Bure Interferometer (PdBI).
Throughout this paper we adopt a $\Lambda$-dominated cosmology: 
$H_0 = 70$~km~s$^{-1}$~Mpc$^{-1}$, $\Omega_\Lambda = 0.7$, and 
$\Omega_m = 0.3$ (Spergel et al.\ 2006).

\section{Observations}

We used the IRAM PdBI to search for the \ci and \water lines in 
APM~08279+5255, redshifted to the 3-millimeter band, on six dates in 
1999 and 2001. 
All of the observations were made in compact configurations of the four 
or five available antennas.
The receivers were tuned to 100.216 GHz 
for the [CI] line, and 113.406 GHz for the H$_2$O line. 
Spectral correlators covered a velocity range of $\sim$1500~km~s$^{-1}$ 
for [CI], and $\sim$1300~km~s$^{-1}$ for H$_2$O.
The phase center was offset by $1.25''$ 
from the CO peak position in D99. 
Baseline lengths ranged from 17 to 81 meters, and the synthesized beam 
sizes were $7\farcs2 \times 6\farcs0$ (position angle 89$^{o}$) for 
the [CI] image, and $5\farcs9 \times 4\farcs4$ (position angle 82$^{o}$) 
for the H$_2$O image.  The nearby quasar 0749+540 was used for complex 
gain calibration.
The flux scale was set using standard sources including MWC349, CRL618, 
3c345.3 and 0923+392, and should be accurate to better than 20\%.
The on-source integration time for the [CI] and H$_2$O images were
equivalent to 7.3 hours and 3.1 hours, respectively, with the full 
six antenna array.
The resulting [CI] image has an rms noise of 0.36~mJy~beam$^{-1}$ 
in a 150~km~s$^{-1}$ channel, and the H$_2$O image has a higher rms noise 
of 0.96~mJy~beam$^{-1}$ in a 130~km~s$^{-1}$ channel, reflecting 
lower atmospheric transmission near the edge of the 3-millimeter window.

\section{Results}

Figure~1 shows images of the [CI] line and continuum emission over 
the full velocity range observed.  The line emission is clearly detected 
in several velocity bins and appears spatially unresolved. 
The position of peak emission is consistent with that found previously 
for the dust continuum, CO lines, and HCN line. 
Figure~2 shows the spectra of the [CI] and H$_2$O lines at this position. 
There is no evidence for significant H$_2$O line emission. 
For the [CI] spectrum, we estimate the continuum level at 100.2~GHz using 
the ``line-free'' channels to be $S_{\rm 100GHz} = 1.18\pm0.18$~mJy.
For the H$_2$O spectrum, we estimate the continuum level at 113.4~GHz using 
all of the channels to be $S_{\rm 113GHz} = 2.43\pm0.48$~mJy.
These values are consistent with the 93.9 GHz continuum flux of 
$1.20\pm0.30$~mJy measured by D99 and a thermal spectrum.
A Gaussian fit to the continuum subtracted \ci spectrum yields a 
peak of $2.20\pm0.51$~mJy,
central velocity $v_0 = 117 \pm 28$~km~s$^{-1}$ (z=$3.9130\pm0.0005$), and 
$\Delta V_{\rm FWHM} = 386 \pm 67$~km~s$^{-1}$. 
The integrated intensity of the \ci line is $0.93 \pm 0.13$~Jy~km~s$^{-1}$, 
which implies a line luminosity
$L'_{\rm CI} = (3.1 \pm 0.4) \times 10^{10}$~K~km~s$^{-1}$~pc$^2$,
following Solomon et al.\ (1992).
Table~1 lists the fitted and derived \ci line parameters.

We place an upper limit on the \water line emission from the noise around 
the mean in the spectrum, assuming 
$\Delta V_{\rm FWHM}$ = 450~km~s$^{-1}$ like the CO, HCN, and [CI] lines.
The 3-$\sigma$ upper limit to the integrated intensity is
$3\cdot(\Delta V_{\rm FWHM}/\Delta V_{\rm chan})^{1/2}\cdot \sigma_{chan} 
= 0.70$~Jy~km~s$^{-1}$, which implies an upper limit to the H$_2$O line 
luminosity $L'_{\rm H_2O} < 1.8 \times 10^{10}$~K~km~s$^{-1}$~pc$^2$.

\section{Discussion}

\subsection{Emission Region and [CI]/CO Luminosity Ratio}

The fitted [CI] line center and width are compatible with previous
observations of the HCN J=5-4 and HCO+ J=5-4 lines (Wagg et al.\ 2005;
Garcia-Burillo et al.\ 2006), though offset by $\sim120$ km~s$^{-1}$ from
the mean redshift determined from the high-J CO lines (D99; Weiss
et al. 2006, in preparation). While the line center difference
between the [CI] and CO is formally significant at the $\sim$4$\sigma$ level,
the true significance is likely lower, since the line profiles from this
complex source are almost certainly not described accurately by a single
Gaussian. Given this uncertainty, the modest signal-to-noise ratios, and
the lack of any resolved velocity structure in the observed line profiles,
we simplify our analysis by assuming the [CI] emission region is cospatial
with the CO line emission region, such that the conditions derived for a
model with a single physical component apply to both species. We stress that
even if all of the observed lines do arise from a single physical component,
the fitted Gaussian centers and widths of lines with different excitation
properties will not necessarily match each other perfectly, given the
complexity of the underlying source structure and optical depth effects.
Data of higher quality and from more transitions will be required to relax
the single component assumption and to justify more sophisticated modeling.

The gas in the single component region is warm and dense, and is likely to be 
a circumnuclear disk of sub-kiloparsec size. The CO emission clearly 
has been magnified by gravitational lensing, and models suggest a 
magnification factor in the range 3 to 20 (D99, Lewis et al.\ 2002), 
or even higher in the models by Egami et al. (2000). 
The exact magnification factor depends critically on the intrinsic size of 
the emitting region.  If the [CI], CO, and HCN emission are truly co-spatial, 
however, then the magnification factors are the same, and line luminosity 
ratios will not suffer any biases due to differential lensing effects.

The luminosity ratio between the [CI] and CO lines is a potential
diagnostic of the gas excitation and carbon chemistry.
To compare the luminosities of [CI] and CO in APM~08279+5255, we use 
the nuclear component of the CO J=1-0 emission (Lewis et al.\ 2002),
which gives $L'_{\rm [CI]} / L'_{\rm CO(1-0)} = 0.23\pm0.06$.  
This value falls within the range $0.2\pm0.2$ found for local 
galaxies (G\'erin \& Phillips 2000), and also the range
$\sim$~0.15~-~0.32 for the other four high-redshift objects that 
have been detected in both [CI] and CO emission.
This ratio does not appear to have any strong dependence on environment, 
either in the local galaxy sample, or at high-redshifts. 
Given the significant uncertainties in the [CI]/CO luminosity ratio
for any individual object, it is not possible to place strong constraints
on the contribution of X-ray irradiation to the molecular gas excitation.

We can estimate the [CI/CO] abundance ratio from the observations, 
adopting the physical conditions derived from multi-transition 
CO observations (Wagg et al. 2005; A. Weiss et al. in preparation)
and a simple radiative transfer model. Using the 
\textit{RADEX}\footnote{http://www.strw.leidenuniv.nl/\~{}moldata/radex.html} 
LVG code (Sch\"{o}ier et al.\ 2005), we find that an abundance ratio, 
[CI/CO]~$\sim$~0.6, reproduces the observed \ci to CO J=1-0 
line luminosity ratio. 
This ratio is similar to that found in the nearby starburst galaxy M82 
([CI/CO]~$\sim$~0.5; Schilke et al.\ 1993, White et al.\ 1994).

\subsection{Neutral Carbon Mass}

The luminosity of the \ci line can be used to derive the neutral carbon mass
(Weiss et al.\ 2005), as
\begin{equation}
M_{[CI]} = 5.706 \times 10^{-4} 
           \frac{Q(T_{ex})}{3}e^{23.6/T_{ex}}L'_{[CI]}~[M_{\odot}], 
\end{equation} 
where $Q(T_{ex}) \sim 1 + 3e^{-T_1/T_{ex}} + 5e^{-T_2/T_{ex}}$ 
($T_1 = 23.6$~K and $T_2 = 62.5$~K are the energies of the $^3$P$_1$ 
and $^3$P$_2$ levels above the ground state) is 
the partition function.
The only high-redshift object for which both the \ci and \cup 
lines have been measured is the ``Cloverleaf'' quasar, where
 $T_{ex} = 30$~K (Weiss et al.\ 2003, 2005). 
Without a measurement of the higher \cup line 
in APM~08279+5255, the value of $T_{ex}$ is uncertain. 
However, Weiss et al.\ (2005) show that for $T_{ex} > 20$~K 
and LTE, the estimated neutral carbon mass is 
insensitive to the assumed excitation temperature. 
Adopting a nominal value of $T_{ex} = 80$~K for the \ci line in 
APM~08279+5255 (from the derived molecular gas and dust temperatures), 
the neutral carbon mass within the nuclear region of APM~08279+5255 
is $M_{\rm CI} = (4.4\pm0.6) m^{-1} \times10^{7}$~M$_{\odot}$,
where $m$ is the lensing magnification factor.
If $T_{ex}$ were $20$~K, then the estimated [CI] mass would be $\sim$30\% 
lower.

\subsection{Molecular Gas Mass}

The luminosity in the CO J=1-0 line is the traditional tracer of molecular gas 
(= H$_2$) mass in the Galaxy and in a variety of extragalactic environments.
  For nearby ultraluminous infrared galaxies 
(ULIRGs; Sanders \& Mirabel 1996; Sanders et al.\ 1988), which are believed 
to be analogues to the currently detectable luminous high-redshift galaxies, 
the empirical scaling factor between CO J=1-0 luminosity and H$_2$ gas mass 
is $\sim$1~M$_{\odot}$~(K~km~s$^{-1}$~pc$^2$)$^{-1}$ (Downes \& Solomon 1998). 
If we apply this factor to the nuclear CO J=1-0 line emission in 
APM~08279+5255 (Papadopoulos et al.\ 2001; Lewis et al.\ 2002), then 
the molecular gas mass implied is $\sim 1.5m^{-1} \times10^{11}$~M$_{\odot}$.

Following Papadopoulos et al.\ (2004a), the small scatter in the observed 
[CI]/CO luminosity ratio suggests that \ci emission may provide an independent 
estimate of molecular gas mass.
Papadopoulos et al.\ (2004a) relate the integrated [CI] line intensity, 
$S_{[CI]}$, to molecular (= H$_2$) gas mass, as 
\begin{equation}
M_{\rm H_2-[CI]} = 1375.8 \frac{D_{l}^2}{1+z}
      \left( \frac{X_{\rm [CI]}}{10^{-5}}  \right )^{-1} 
    \left ( \frac{A_{10}}{10^{-7}s^{-1}} \right )^{-1} 
   Q_{10}^{-1}\frac{S_{\rm [CI]}}{\rm [Jy~km~s^{-1}]}~[M_{\odot}],
\end{equation} 
where $X_{\rm CI}$ is the [CI]-to-H$_2$ abundance ratio
(we assume $X_{\rm CI} = 3 \times 10^{-5}$), $A_{10}$ is the Einstein 
A-coefficient ($A_{10} = 7.93 \times 10^{-8}$~s$^{-1}$), and $Q_{10}$ is 
the excitation factor which depends on the kinetic temperature, and 
density of the gas (we adopt $Q_{10} = 0.5$). The cosmology dependence is
 included through the luminosity distance, $D_l$.
Using the observed integrated intensity of the \ci line in APM~08279+5255 
gives $M_{\rm H_2 - [CI]} = (2.7\pm0.4)m^{-1} \times 10^{11}$~M$_{\odot}$,
in good agreement with the molecular gas mass estimated from the CO J=1-0 line.

Table~\ref{table2} presents the molecular gas mass estimates derived
from [CI] and from CO for the five high-redshift objects where both lines
are detected, adjusted to a common cosmology.
The \ci line integrated intensities for IRAS~F10214, SMM~J14011+0252, 
and H1413+117 are taken from Weiss et al.\ (2005), and for
PSS~2322+1944 from Pety et al.\ (2005).
The CO J=1-0 line has been detected only in APM~08279+5255 and
PSS~2322+1944 (Carilli et al.\ 2002). For the other objects, which are
not detected in the CO J=1-0 line, we use observations of 
the CO J=3-2 line (IRAS~F10214; Solomon et al.\ 1992, H1413+117; 
Weiss et al.\ 2003, SMM~J14011+0252; Downes \& Solomon 2003),
which is approximately equal to the luminosity in the CO J=1-0 line 
at high-redshifts for warm, dense gas (Solomon et al.\ 1992).
Though the uncertainties are large, there is very good agreement between 
the molecular gas mass estimates from the two tracers.

\subsection{H$_2$O}

We did not detect any significant \water line emission in APM~08279+5255.
We can estimate the H$_2$O line luminosity expected for the circumnuclear 
region for a nominal water abundance, again adopting the previously derived 
physical conditions for the region.
In Galactic molecular clouds, a typical ortho-H$_2$O abundance is 
$4.5\times10^{-9}$ relative to H$_2$ (Snell et al. 2000; Ashby et al. 2000).
For $T_{kin}$ = 80~K, $n_{H_2} \sim 40,000$~cm$^{-3}$, an LVG calculation
gives $N({\rm H_2O})/\triangle v = 
8.0 \times 10^{12}$~cm$^{-2}$(km~s$^{-1}$)$^{-1}$,
$L'_{\rm H_2O} = 1.1 \times 10^{7}$~K~km~s$^{-1}$~pc$^2$, 
more than two orders of magnitude below the upper limit.
In strongly shocked regions, the H$_2$O abundance has been observed to be 
enhanced by more than an order of magnitude, but even if that were the case, 
calculations show that the \water line emission would remain well below 
the achieved detection threshold. 

\section{Summary}

We detected \ci line emission in the ultraluminous quasar
APM~08279+5255 at z=3.91. The \ci line width and center are 
similar to those of previously detected millimeter CO and HCN lines.
Estimates of the molecular gas mass based on the \ci line and 
based on CO lines yield similar results, and there is no 
evidence for any substantial contribution from a primarily 
atomic medium. Though the observational uncertainties remain large,
this seems to be the case for the five high-redshift sources 
with reported detections of both CO and [CI] emission lines.
The \ci line may be a valuable alternative probe of molecular gas 
mass in systems lacking high excitation CO line emission,
where the molecular medium is more diffuse and cooler 
($T_{kin} \la 20$~K,and $n_{H_2} \la 10^4$~cm$^{-3}$; 
Papadopoulos et al.\ 2004a).  The Atacama Large Millimeter Array 
will greatly expand the number of high-redshift sources accessible 
in [CI] and CO emission, and will have the capability to spatially 
resolve the emission to show directly the extent of the gas.

\section{Acknowledgments}

We thank the IRAM PdBI staff for carrying out these observations. 
J.W. is grateful to the SAO for support through a predoctoral student 
fellowship and the Department of Astrophysics at INAOE for a graduate student
scholarship. This work is partially supported by CONACYT grant 39953-F. 
J.W. thanks S\'ebastien Muller for helpful tips on PdBI data reduction, 
and also Padeli Papadopoulos and Matthew Ashby for discussions about 
[CI] and H$_2$O excitation and emission at high-redshift. We thank
 the referee for a thorough reading of the submitted manuscript
and helpful suggestions.

\clearpage

\begin{deluxetable}{lc}
\tablecaption{\ci line parameters for APM~08279+5255.
\label{table1}}
\tablewidth{0pt}
\tablehead{\colhead{} & \colhead{}}
\startdata
\ci peak:        & $2.20\pm0.51$~mJy \\
\ci $\Delta V_{\rm FWHM}$: &  $386\pm67$~km~s$^{-1}$  \\
\ci $v_0$\tablenotemark{a}: & $117\pm28$~km~s$^{-1}$ \\
$S_{[CI]}$: &  $0.93\pm0.13$~Jy~km~s$^{-1}$  \\
$L'_{[CI]}$\tablenotemark{b}: & $3.1\pm0.4 \times 10^{10}$~K~km~s$^{-1}$~pc$^2$  \\
\enddata            
\tablenotetext{a}{velocity with respect to z=3.911, determined from CO 
lines.}
\tablenotetext{b}{The \ci luminosity is not corrected for lensing
 magnification.}
\end{deluxetable}

\clearpage

\begin{deluxetable}{lccccc}
\tablecaption{Masses of neutral carbon and H$_2$ in high-redshift objects.
\label{table2}}
\tablewidth{0pt}
\tablehead{\colhead{Object}  & \colhead{$z_{CO}$\tablenotemark{a}}  & 
\colhead{M$_{[CI]}$}  &
\colhead{M$_{H_2}$(CO)} &
\colhead{M$_{H_2}$([CI])} &
\colhead{Refs.} \\
\colhead{} & \colhead{} & 
\colhead{[M$_{\odot}$/10$^7$]}  &
\colhead{[M$_{\odot}$/10$^{10}$]} &
\colhead{[M$_{\odot}$/10$^{10}$]} &
\colhead{CO,[CI]}
} 
\startdata
IRAS F10214  & 2.29 & (2.7$\pm$0.3)$m^{-1}$ & (11.3$\pm$2.5)$m^{-1}$ &  (18.8$\pm$2.4)$m^{-1}$ & 1,2 \\
H1413+117 & 2.56 & (8.1$\pm$1.2)$m^{-1}$ & (44.3$\pm$0.7)$m^{-1}$ &  (55.8$\pm$8.6)$m^{-1}$  & 3,2 \\
SMM~J14011+0252 & 2.57 & (3.7$\pm$0.6)$m^{-1}$ & (9.5$\pm$1.0)$m^{-1}$ &  (25.9$\pm$4.3)$m^{-1}$ & 4,2 \\
\textbf{APM~08279+5255} & 3.91 & (4.4$\pm$0.6)$m^{-1}$ & (13.4$\pm$3.0)$m^{-1}$ & (26.7$\pm$3.7)$m^{-1}$  & 5,6\\
PSS 2322+1944 & 4.12 & (3.8$\pm$0.6)$m^{-1}$ & (12.5$\pm$5.3)$m^{-1}$ &  (25.2$\pm$3.7)$m^{-1}$ &  7,8 \\
\hline
\enddata 
\tablenotetext{a}{Redshift of CO emission.}   
\tablecomments{$m$ is the gravitational lensing magnification factor \\
References -- (1) Solomon et al. 1992; (2) Weiss et al.\ 2005; (3) Weiss et al.\ 2003;
 (4) Downes \& Solomon 2003; (5) Lewis et al.\ 2002; (6) This paper; 
(7) Carilli et al.\ 2002; (8) Pety et al.\ 2005
}
\end{deluxetable}

\clearpage

\begin{figure}
\centering
\includegraphics[height=6.0in,angle=270]{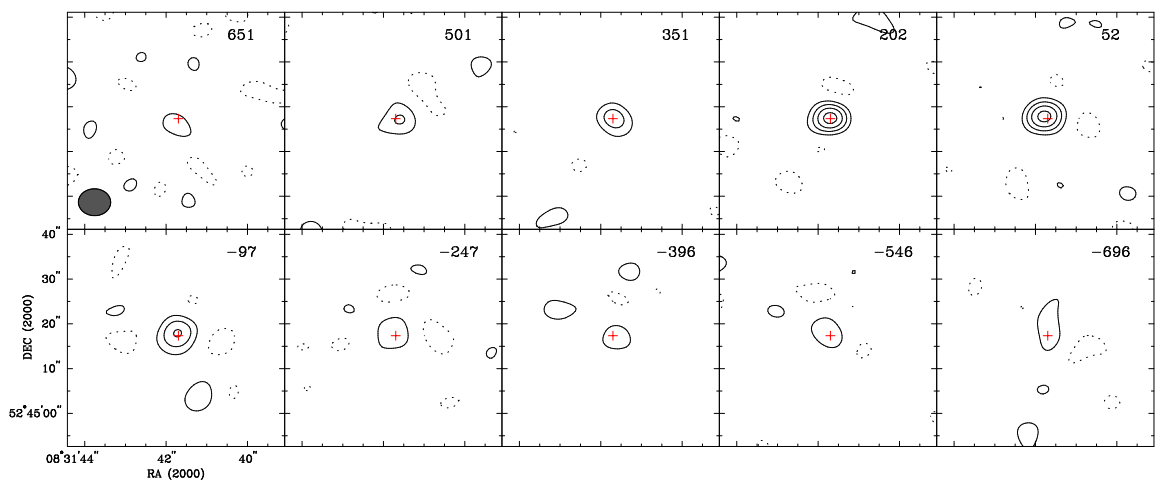}
\caption{
Images of the \ci and 100.2 GHz continuum emission in APM~08279+5255
over the full observed velocity range. 
The contour intervals are -2, 2, 4, 6 and 8$\times\sigma$ 
(0.36~mJy~beam$^{-1}$). 
The cross marks the position of the CO peak position from D99, which is
offset ($1\farcs19$, $-0\farcs36$) from the phase center 
(08$^h$31$^m$41$^s$.57, 52$^o$45$^m$17$^s$.7). The filled ellipse in the
top left panel shows the 
$7\farcs2 \times 6\farcs0$ PA 89.0$^{\circ}$ synthesized beam. 
The velocities indicated in the upper right corner in each panel are 
relative to z=3.911, determined from CO observations.
\label{fig:ci_chan}.}
\end{figure}

\clearpage

\begin{figure}
\centering
\includegraphics[width=5in]{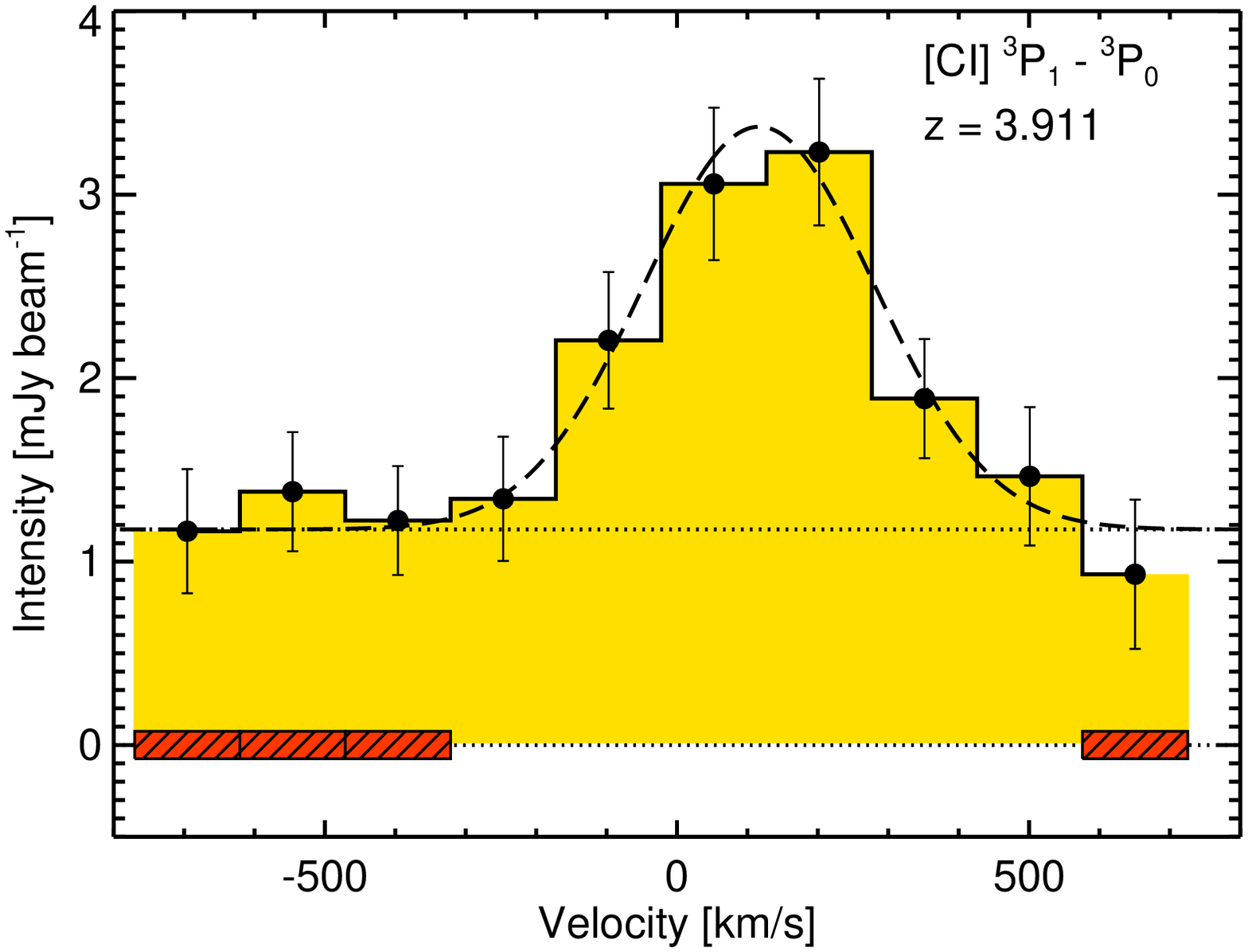}
\includegraphics[width=5in]{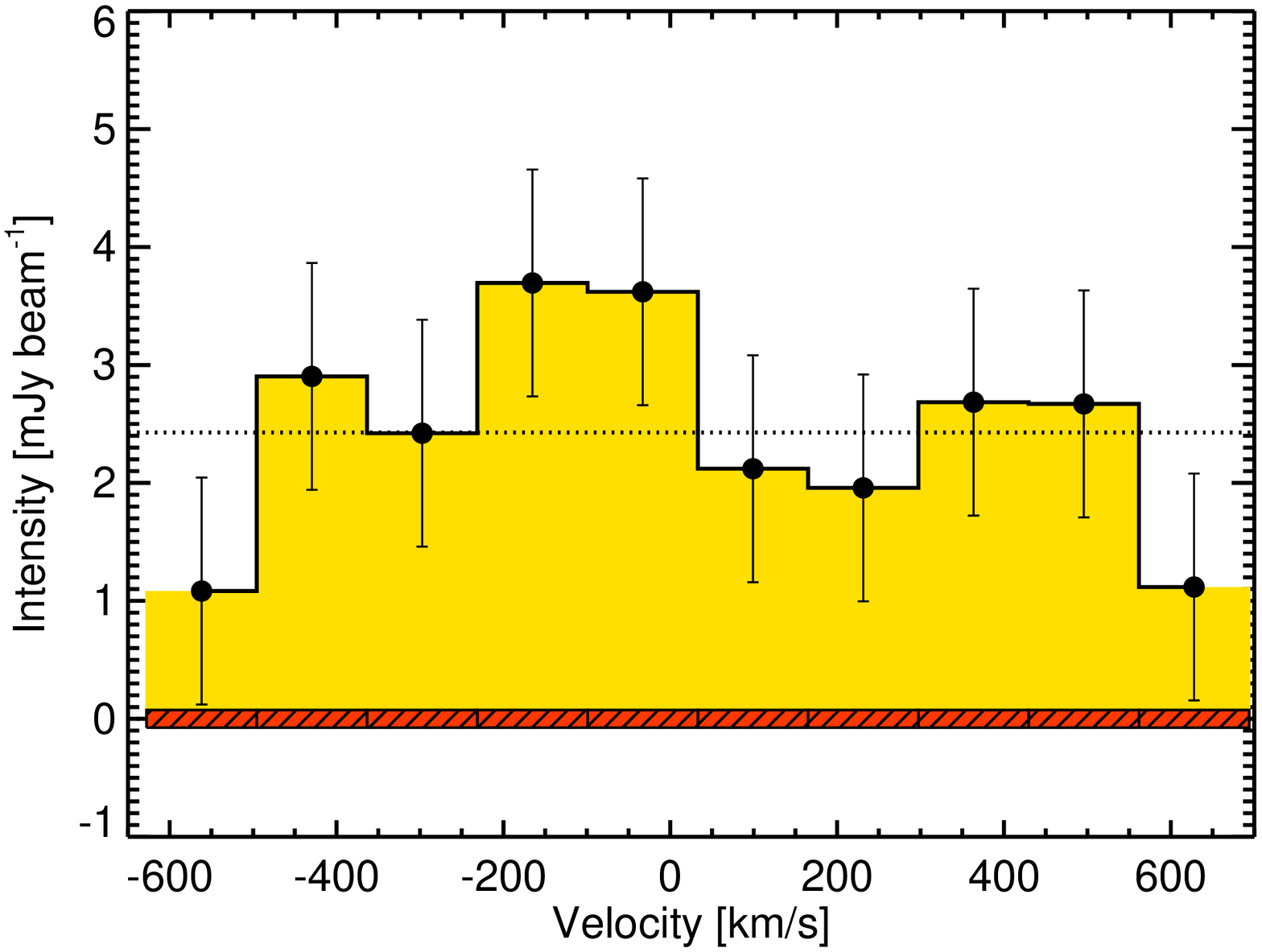}
\caption{ Spectra of \ci (\textit{top}) and \water (\textit{bottom}) at 
the position of peak continuum intensity. 
Dotted lines show the fitted continuum levels, calculated from the intensity 
in the channels marked by the hatched regions. The dashed line shows 
a Gaussian fit to the \ci emission. 
\label{fig:ci_spec}}
\end{figure}


\begin{references}

\reference{ashby00} Ashby, M.~L.~N., et al.\ 2000, ApJ, 539, L115

\reference{barvainis97} Barvainis, R., Maloney, P., Antonucci, R., 
           Alloin, D., 1997, ApJ, 484, 695

\reference{carilli02} Carilli C.~L., Cox P., Bertoldi F., Menten K.~M., 
Omont A., Djorgovski S.~G., Petric A., Beelen A., Isaak K.~G., McMahon R.~G.
2002, ApJ, 575, 145

\reference{1994A&A...287..716C} Casoli F., Gerin M., Encrenaz P.~J., 
Combes F., 1994, A\&A, 287, 716 

\reference{combes_wiklind} Combes, F. \& Wiklind, T. 1997, ApJ 486, L79 

\reference{downes98} Downes, D. \& Solomon, P.M. 1998, ApJ, 507, 615

\reference{1999ApJ...513L...1D} Downes D., Neri R., Wiklind T., 
          Wilner D.\ J., Shaver P.\ A. 1999, ApJ,  513, L1 (D99)

\reference{DS2003} Downes, D. \& Solomon, P.M. 2003, ApJ, 528, 37

\reference{egami00} Egami E., Neugebauer G., Soifer B.\ T., Matthews K., Ressler M., Becklin E.\ E., Murphy T.\ W., Dale D.\ A., 2000, ApJ,  535, 561


\reference{1999PASP..111..946E} Ellison S.~L., Lewis G.~F., Pettini M., Sargent W.~L.~W., Chaffee F.~H., Foltz C.~B., Rauch M., Irwin M.~J., 1999, PASP, 111, 946 

\reference{1993A&A...273L..19E} Encrenaz P.~J., Combes F., Casoli F., 
Gerin M., Pagani L., Horellou C., Gac C., 1993, A\&A, 273, L19 
 
\reference{2002ApJ...567...37G} Gallagher S.~C., Brandt W.~N., Chartas G., Garmire G.~P., 2002, ApJ, 567, 37 

\reference{2006astro.ph..5656G} Garcia-Burillo S., et al., 2006, astro-ph/0605656 

\reference{GP98} G\'erin M. \& Phillips T.G. 1998, ApJ, 509, L17

\reference{GP00} G\'erin M. \& Phillips, T.G. 2000, ApJ, 537, 644


\reference{ibata99} Ibata R.\ A., Lewis G.\ F., Irwin M.\ J., Leh{\'a}r J., Totten E.\ J., 1999, AJ,  118, 1922 


\reference{irwin98} Irwin M.\ J., Ibata R.\ A., Lewis G.\ F., Totten E.\ J., 1998, ApJ,  505, 529  

\reference{lewis98} Lewis, G.F., Chapman, S.C., Ibata, R.A., Irwin, M.J., \& Totten, E.J. 1998, ApJ, 505, L1 

\reference{lewis02} Lewis, G.F., Carilli, C., Papadopoulos, P., Ivison, R.J., 2002, MNRAS, 330, L15 

\reference{maloney96} Maloney, P.R., Hollenbach, D.J., Tielens, G.G.M. 
             1996, ApJ, 466, 561

\reference{papadopoulos01} Papadopoulos, P., Ivison, R.J., Carilli, C., Lewis., G. 2001, Nature, 409, 58


\reference{papadop04a} Papadopoulos, P.P., Thi, W.-F., Viti, S. 2004a,
                      MNRAS, 351, 147

\reference{papadop04b} Papadopoulos, P.P. \& Greve, T.R. 2004b, ApJ, 615, L29

\reference{papadop05} Papadopoulos P.~P., 2005, ApJ, 623, 763 
 
\reference{pety04} Pety J., Beelen A., Cox P., Downes D., Omont A., 
      Bertoldi F., Carilli C.~L. 2005, A\&A, 428, L21
 


\reference{sanders88} Sanders, D.B., Soifer, B.T., Elias, J.H., Madore, B.F., Matthews, K., Neugebauer, G., \& Scoville, N.Z. 1988, ApJ, 325, 74

\reference{sanders96} Sanders, D.B. \& Mirabel, I.F. 1996, ARA\&A, 34, 749

\reference{schilke93} Schilke, P., Carlstrom, J.E., Keene, J., Phillips, T.G., 1993, ApJ, 417, L67

\reference{radex} Sch\"{o}ier, F.L., van der Tak, F.F.S., van Dishoek, E.F., Black, J.H., 2005, A\&A, 432, 369

\reference{snell00} Snell, R.~L., et al.\ 2000, ApJ, 539, L101

\reference{2004ApJS..154..151S} Soifer B.~T., et al.\ 2004, ApJS, 154, 151

\reference{solomon92b} Solomon, P.M., Downes, D., Radford, S.J.E., 1992, ApJ, 398, L29 

\reference{2006astro.ph..3449S} Spergel D.~N., et al., 2006, astro-ph/0603449 

\reference{apmhcn} Wagg J., Wilner D.\ J., Neri R., Downes D., 
          Wiklind T. 2005, ApJ, 634, L13 

\reference{2003A&A...409L..41W} Weiss A., Henkel C., Downes, D., Walter F. 2003, A\&A, 409, L41 

\reference{2005A&A...429L..25W} Weiss A., Downes D., Henkel C., Walter F. 2005, A\&A, 429, L25 

\reference{white94} White, G.J., Ellison, B., Claude, S., Dent, W.R.F., Matheson, D.N., 1994, A\&A, 284, L23 


\end{references}
\end{document}